\documentclass[tightenlines,aps,twocolumn,showpacs,floatfix]{revtex4}

\usepackage{amsmath,amsfonts,graphicx,bm}
\begin{document}      
%

%


\preprint{\vbox{\baselineskip=14pt%
   \rightline{???-HEP-???}      
}}      
\title{
\[ \vspace{-2.1cm} \]
\noindent\hfill\hbox{\rm\normalsize \textsf{SLAC-PUB-9652}} \vskip 1pt
\noindent\hfill\hbox{\rm\normalsize \textsf{UH-511-1018-03}} \vskip 1pt
\noindent\hfill\hbox{\rm\normalsize \textsf{hep-ph/0302014}} \vskip 10pt
The self-energy of improved staggered quarks
}      
\author{Thomas Becher$\phantom{}^a$ and Kirill Melnikov$\phantom{}^b$}      
    \affiliation{$\phantom{}^a$Stanford Linear Accelerator Center, Stanford University,
Stanford, CA 94309, U.S.A. \\
$\phantom{}^b$Department of Physics and Astronomy, Univ.~of Hawaii, 2505 Correa Rd., Honolulu, HI 96822, U.S.A.}        

\begin{abstract}    
We calculate the fermion self-energy at ${\cal O}(\alpha_s)$ for the
Symanzik improved staggered fermion action used in recent lattice
simulations by the MILC collaboration.  We demonstrate that the
algebraic approach to lattice perturbation theory, suggested by us
recently \cite{Becher:2002if}, is a powerful tool also for
improved lattice actions.
\end{abstract}      
\maketitle     

\section{Introduction}

In a recent paper \cite{Becher:2002if}, we have shown that methods
familiar from continuum perturbation theory can be used to perform
calculations in lattice perturbation theory analytically, except for
the numerical evaluation of a small number of lattice
``master''-integrals. In short, the calculation proceeds as follows:
in a first step, a given loop integral is expanded around the
continuum limit using the technique of ``asymptotic expansions'' for
Feynman diagrams, familiar \cite{Smirnov:pj} from continuum
perturbation theory.  The expansion splits the original integral into
continuum integrals and a set of lattice tadpole integrals. These
tadpole integrals are independent of external momenta and masses and
are interrelated through a set of algebraic identities. Using these
relations, the tadpole integrals can be expressed through a small
number of master integrals which are evaluated numerically. In
\cite{Becher:2002if}, we have illustrated this strategy by evaluating
the gluon and fermion one-loop self-energies with the standard Wilson
action.

We have also asserted that the above techniques are not limited to a
particular lattice action or to one loop. While correct in principle,
in practice this could turn out to be of little use, given the
limitations of available computing power. The goal of the present
paper is to investigate if the method proposed in \cite{Becher:2002if}
remains useful for improved lattice actions which are employed in many
lattice simulations.  In what follows, we perform an analytic one loop
calculation with the lattice QCD action used by the MILC collaboration
in their recent simulations with three dynamical quarks
\cite{Bernard:2001av}. This is a one-loop Symanzik improved gluon
action \cite{Luscher:1985zq} coupled to improved Kogut-Susskind
fermions \cite{Kogut:ag,Lepage:1998vj}. This action is accurate up to
errors of $O(a^2\alpha_s,a^4)$.  We find that the amount of computer algebra
needed for perturbative calculations with such an action is
significantly larger than for the simple Wilson action, but they
nevertheless turn out to be feasible.

Let us stress an important difference between the purely numerical
approach used e.g.~in \cite{Nobes:2001tf} and the algebraic method \cite{Becher:2002if} which we use in this paper.  In order to
make use of all algebraic relations, not only the integrals that
appear explicitly in the expansion of a given graph, but the entire
class of lattice tadpole integrals relevant for a given lattice action
needs to be considered. For the present calculation we had to produce
a database of roughly 30'000 tadpole integrals that are all expressed
in terms of 16 master integrals. This took about a week of CPU time
running Maple on a desktop computer. Less than 2000 of those integrals
actually occurred in our diagrams. On the other hand, once the
database is set up, it can be used for any one-loop calculation
involving staggered fermions. In this paper we deal with the simplest
example of the quark self-energy and derive the relation between the
bare and the pole mass of the quark at order ${\cal O}(\alpha_s)$.  We
plan to return to other applications
in the near future.

This paper is organized as follows. We start by summarizing the method
proposed in \cite{Becher:2002if} and by pointing out some
peculiarities in its application to staggered fermions.  The presence
of the fermion doublers manifests itself in the appearance of multiple
``soft'' regions. We then explain our treatment of the improvement
terms in the action.  In our calculation we expand the improved gluon
and fermion propagators in a geometric series around the standard
propagators. Since loop integrals are analytic in the improvement
terms, this expansion can be performed at the level of the
integrand. As a matter of principle, it can be carried to any
level of precision desired. Finally, we present the result for the
fermion self-energy and conclude.

\section{Asymptotic expansion around the continuum limit}

In this Section we briefly describe the main idea of our approach to
lattice perturbation theory \cite{Becher:2002if}.  We are interested
in the calculation of quantities in the continuum limit, where the
particle masses and external momenta are much smaller than the inverse
lattice spacing. It is possible to construct the expansion of a
lattice integral around the continuum limit by expanding the {\em
integrand}, at the expense of introducing an additional (analytic)
regulator.

It turns out to be convenient to map \footnote{We use
dimensionless units throughout the paper, so that the lattice spacing
is $a=1$.}  the integration region $-\pi < k_\mu < \pi$ to infinite
volume, by changing the integration variables to
\begin{equation}   
\eta_\mu = \tan(k_\mu/2).
\label{beq}
\end{equation}
Furthermore, we introduce an intermediate regulator by adding a small
quantity $\delta$ to the power of one of the propagators. Keeping the
regularization parameter $\delta$ non-zero, we then expand the
integrand of the lattice loop integral in the so-called soft and hard
regions. The original integral is given by the sum of the soft and
hard contributions; in this sum all the terms that are singular in the
limit $\delta \to 0$ cancel out. The soft and hard contributions are
calculated as follows:
\begin{itemize}
\item Soft: assume that all the components of the
loop momentum $\eta$  are small, i.e. they 
comparable to masses $m$ and external momenta $p$ in the problem, 
$\eta_i \sim m,~p \ll 1 $.
Perform the Taylor expansion of  the integrand in
the small quantities $\eta_i$ and $m,p$. 
The expansion coefficients in this region
are standard continuum one loop integrals, regularized analytically.
No restriction on the integration region is
introduced.
\item Hard: assume that all the components of the loop momentum are large,
$\eta_i \sim 1 \gg m,p$ and Taylor expand the integrand in $m,p$. 
The expansion coefficients are massless lattice tadpole integrals.
\end{itemize}
As in dimensional regularization, scaleless integrals are set to
zero.

While the soft part is given by continuum integrals, the hard part is
given by lattice tadpole integrals that do not depend on the process
under consideration.  Intuitively, this is the case because the hard
part arises from the integration region where the loop momentum is
comparable to the inverse lattice spacing and therefore much larger
than the external momenta. The analytic regularization which we use in
what follows, turns out to be very efficient in separating soft and
hard momenta modes and allows us to treat them separately.  It is
important to realize that since the integrals occurring in the hard
part are process independent, they can be calculated once and for all
for a given lattice action.

In any realistic calculation, many lattice tadpole integrals
appear. Fortunately, not all of them are independent. Relations
between the integrals are obtained after observing that the integral of a
total derivative with respect to one of the variables $\eta_\mu$
vanishes (there are no surface terms, thanks to analytic
regularization). By explicitly computing the derivative, one
finds the algebraic relations (recurrence relations) between the
lattice tadpole integrals.  It is possible to solve these
relations using the Gaussian elimination method and express all the lattice
tadpole integrals through a small number of master integrals. For more details
we refer to \cite{Becher:2002if}.

The strategy outlined above does not rely on the specific form of the
propagator. In \cite{Becher:2002if}, we have applied the technique to
HQET and to QCD with Wilson fermions and calculated a number of QCD
one-loop self-energies.  We now show how to use the method to
calculate loop integrals for staggered fermions. The staggered fermion
action is obtained by reducing the number of components of the fermion
field after spin diagonalizing the naive lattice fermion action. In
perturbative calculations it is convenient to work with the naive
fermion action and perform the staggering only at the end of the
calculation.

The fermion doubling inherent in the
naive discretization of the fermion action manifests itself in the
appearance of multiple soft regions. The sixteen zeros of the inverse
propagator of naive fermions,
\begin{equation}
D_F=\sum_\mu \frac{1}{4}\sin^2 k_\mu
=\sum_\mu \frac{\eta_\mu^2}{(1+\eta_\mu^2)^2}\, ,
\end{equation}
in the Brillouin zone give rise to sixteen propagating
fermions. Correspondingly loop integrals which involve naive fermion
propagators have sixteen soft regions. One of those is the region
where all components of the loop momentum $\eta_i$ are small. The
fifteen additional doubler contributions arise after transforming one
or several components of the integration momentum as
$\eta_i\rightarrow 1/\eta_i$ and then expanding the resulting
integrand around small $\eta_i$. For purely fermionic integrals, the
sixteen soft contributions are all equal. In integrals with both boson
and fermion propagators, this is not the case:
since the boson propagator is far off-shell in all 
corners of the Brillouin zone, 
the doubler contributions are different from the genuine soft contribution 
where all the momenta are small compared to inverse lattice spacing.

The contribution of the hard part is obtained as usual, by expanding the loop
integral in external momenta and particle masses. The lattice tadpole integrals
that occur in the case at hand are
\begin{equation}
H(\{a_i\},n,m)=\prod_{i=1}^4\int \frac{d\eta_i}{(1+\eta_i^2)^{a_i}}\,\frac{1}{D_B^n}\,\frac{1}{D_F^{m+\delta}},
\label{eqH}
\end{equation}
where $D_B$ is the standard bosonic propagator expressed through 
variables $\eta$.
Let us stress that the regulator {\it has} to be on the fermion propagator 
in order to regulate both the 
singularities at $\eta_\mu=0$ {\em and} $\eta_\mu=\infty$. 
The integration-by-parts and partial fractioning relations for this class of integrals are
\begin{eqnarray}
0&=&\bigg\{1 + 2\, a_i\,({\bm a}_i-1) + 2\,{\bm a}_i\, ({\bm a}_i-1)\, n\,{\bm n} \nonumber\\
&&+ 2\,{\bm a}_i\, (\,{\bm a}_i-1)\, (2\,{\bm a}_i-1)\, (m+\delta) \,{\bm m}\bigg\}\,H\, ,  \nonumber\\ 
&& \nonumber\\ 
0&=&\bigg\{1-  \sum_{i=1}^{d}\,\,{\bm a}_i\,(1 - \,{\bm a}_i)\,{\bm m}\bigg\}\,H \, , \\ 
0&=&\bigg\{1 - \sum_{i=1}^{d} (1 - \,{\bm a}_i)\,{\bm n}\bigg\}\,H\, .\nonumber
\end{eqnarray}
The operator ${\bm a}_i$ raises the index $a_i$ by one unit, etc.  The
above relations are sufficient to reduce all integrals $H$ to sixteen
master integrals. These can be chosen to be convergent (in the limit
$\delta\rightarrow 0$) and are calculated numerically. An efficient
method for their evaluation is described in \cite{Becher:2002if}.

\section{Improved actions}

The lattice artifacts arising in calculations with the simplest
discretization of the QCD action turn out to be uncomfortably large,
making it difficult to extract physical quantities from simulations
with currently accessible lattice spacings. For this reason, many
simulations are done with improved actions, which include higher
dimensional operators whose coefficients are tuned in such a way as to
reduce the cutoff effects arising at finite lattice spacing.

In principle one could apply the procedure for perturbative 
calculations described above  directly to improved
actions. The higher derivative terms encountered in these actions
complicate the calculation in two ways: (1) The recursion relations
for the improved propagators are much more complicated than for the
standard case which might lead to problems with finding their 
solutions and, also, to a larger number of master integrals;
(2) The derivative interactions make it necessary to calculate
hard integrals with more complicated ``tensor'' structure, i.e. higher
powers of $a_i$ in Eq.(\ref{eqH}).

In the following we avoid the first complication by expanding the
improved propagators around the standard ones. In the soft region,
where the momentum flowing through the propagators is small, this
expansion is built into our formalism since  for small momenta the higher
dimensional operators are suppressed by powers of the lattice
spacing. For large momentum, on the other hand, these operators are
suppressed only numerically, but not parametrically. As we will 
show this suffices to get the result with reasonable precision.

The disadvantage of expanding the propagators is that very high
powers of lattice derivatives appear. 
For example, the quark gluon coupling of the
Symanzik improved staggered quark action involves terms with up to
six derivatives. The usual fermion self-energy graph therefore involves 12
powers of the lattice momentum $\widehat k_\mu$ just from the 
vertex plus additional powers arising from the expansion of 
the two propagators.  This is why we were forced to  compute 
approximately thirty thousand lattice tadpole integrals to have the 
full reduction of the integrals that appear in the calculation of the 
fermion self-energy.

\subsection{Improved staggered fermions}
The naive discretization of the free fermion action can be improved by
adding a third lattice derivative term to the action:
\begin{equation}
{\cal L}={\bar\psi}_x\left[ \sum_\mu\Delta_\mu (1+\frac{b}{6}\Delta^2_\mu)\gamma_\mu+m\right] \psi_x\label{naive}\,.
\end{equation}
The associated fermion propagator
\begin{equation}
D_F^{-1}(k)=i \sum_\mu \gamma_\mu \sin k_\mu\,(1+\frac{b}{6}\sin^2 k_\mu)+m\, 
\end{equation}
reproduces, for $b=1$,  the {\em free} continuum propagator 
up to terms suppressed as $O(k^4)$ at low energies. Note that the third order
derivative in the above action not only improves the
fermion propagator at the origin of the Brillouin zone, but also in 
all of its corners where some components of the momentum are
$\pi$ and doublers appear.

From the above expression it is obvious that the improvement term
proportional to $b$ is numerically suppressed
over the entire momentum range
$-\pi<k<\pi$. For the propagator at the
origin
\begin{multline}
\int_{-\pi}^\pi\frac{d^4 k}{(2\pi)^4}\,\frac{1}{\sum_\mu \sin k_\mu^2\,(1+b\sin^2 k_\mu )^2}=\\
0.6197336 - 0.1267555\,b + 0.0214647\,b^2 \\
- 0.00331\,b^3 + 0.0004803\,b^4 + \dots
\end{multline}
the expansion in $b$ is roughly a geometric series with an expansion parameter of $1/6$.

The above action is no longer free of $O(a^2)$ lattice artifacts,
once a gauge field is coupled to the fermion. The standard
discretization of the covariant derivative
\begin{align}
D_\mu \psi_x&=\frac12 \left[U_{x,\mu}\psi_{n+\hat\mu}-U_{x-\hat\mu,\mu}\psi_{n-\hat\mu}\right], \nonumber\\
\label{Ulink}
U_{x,\mu}&=e^{-i g A_{x,\mu}} 
\end{align}
leads to ``flavor''-changing interactions, in which a fermion emits gluon with momentum of order $\pi$ and 
scatters into one of its fifteen partners. The Feynman rule for the quark-gluon coupling of naive fermions 
\begin{equation*}
i\,g \, (2\pi)^4\,\delta^4(k_q-k_{\bar q}+k_g)\,\sum_\mu \gamma_\mu\,\cos(\frac{k_q^\mu+k_{\bar q}^\mu}{2})
\end{equation*}
shows that the flavor-changing and the flavor-neutral interaction have
the same strength. However, in flavor changing interactions momentum of the
order of the inverse lattice spacing is transferred to the gluon,
taking it far off-shell and thus suppressing the interaction by
$O(a^2)$.

To remove the flavor changing interactions at tree level, Lepage suggested
\cite{Lepage:1998vj} to replace the link field $U_\mu(x)$ in the
covariant derivative by a ``fat'' link
\begin{equation}\label{Vlink}
V'_{x,\mu} := 
\prod_{\nu\ne\mu}\left.\left(1 + \frac{\Delta^{(2)}_\nu}{4}\right)
         \right|_{\rm symm.} U_{x,\mu}- \frac14 \sum_{\nu\ne\mu}
(\nabla_\rho)^2 U_{x,\mu}\,.
\end{equation}
The derivative  interactions in the first term of Eq.(\ref{Vlink})
have been chosen such that the ``flavor''-changing interactions get suppressed,
while the second term improves the ``flavor''-conserving part. Note
that the replacement $U_\mu(x)\rightarrow V_\mu'(x)$ is only done for
the first derivative term in Eq. (\ref{naive}).

\subsection{Improved gluon propagator}
The tree level improved gauge action has the form
\begin{equation}
{\cal L}= -\beta\,\sum_{x,\mu>\nu} \{ c_1 P_{\mu\nu}(x) + c_2 (R_{\mu\nu}+R_{\nu\mu}) \}, \label{eq:action}
\end{equation}
where $\beta = 6/g_0^2$, $c_1 = 5/3$ and $c_2 = -1/12$.
The plaquette $P_{\mu\nu}$ is the $1\times 1$ Wilson-loop in the
$\mu\nu$-plane and the rectangle $R_{\mu\nu}$ is the loop with two steps
in the $\mu$ and one in the $\nu$ direction.  Parameterizing
the link between the lattice site $n$ and $n+\hat\mu$ as
\begin{align*}
A_{x,\mu}&=\int_p\,e^{i(n+\hat\mu/2)p}\, A_\mu(p)\,,
\end{align*}
the quadratic part of the gluon Lagrangian in Fourier space is
\begin{equation}
{\cal L}_{A^2}= A_\mu(-p) \left[ D_{\mu\nu}+E_{\mu\nu}\right] A_\nu(p)
\end{equation}
with ($\hat p_\mu = 2 \sin p_\mu/2$)
\begin{align}
D_{\mu\nu}&= {\widehat p^2}\,g_{\mu\nu}-\frac{\xi}{1+\xi}\, {\widehat p}_\mu {\widehat p}_\nu, \\
E_{\mu\nu}&=\frac{1}{12}\left[(\mbox{$\sum_\mu$} {\widehat p}_\mu^4 + {\widehat p^2} {\widehat p}_\mu^2)\,g_{\mu\nu}-({\widehat p}_\mu^3\, {\widehat p}_\nu +{\widehat p}_\mu\, {\widehat p}_\nu^3)\right] \,.\nonumber
\end{align}
The first term $D_{\mu\nu}$ corresponds to
the usual plaquette action obtained by setting 
$c_1=1$, $c_2=0$ in Eq.(\ref{eq:action}). 
The corresponding propagator is\footnote{With this definition, 
$\xi=0$ is Feynman gauge and $\xi=-1$ corresponds to Landau gauge.} 
\begin{equation}
G_{\mu\nu}=D^{-1}_{\mu\nu}=\frac{g_{\mu\nu}}{\widehat p^2}+\xi\frac{\widehat p_\mu\,\widehat p_\nu}{(\widehat p^2)^2} \,.
\end{equation}
We treat $E$ as a perturbation and expand the improved propagator as
\begin{equation*}
G^{\rm Imp}=G-b\;G\cdot E \cdot G+b^2\;G\cdot E \cdot G\cdot E \cdot G - \dots
\end{equation*}
The hat variables vary between $-2<{\widehat p_\mu}<2$. The
effect of the improvement is largest in the corners of the Brillouin
zone, where the above expansion is a  geometric series with expansion
parameter $2/3$. For the loop graphs calculated below, the expansion parameter turns out to be around $1/2$. 
To give an example, consider the expansion of the $\mu\mu$-component of the 
gluon propagator at the origin.  It has the following form:
\begin{multline}
G^{\rm Imp}_{\mu\mu}(x=0)= 0.1549334\,(1+\xi/4) - 0.0366778\,b \\
+ 0.014347\,b^2 - 0.0063397\,b^3 + 0.003045\,b^4
\\ - 0.0015511\,b^5 + 0.0008239\,b^6+\dots
\label{eqprop}
\end{multline}
All given digits are significant. To a good approximation, the expansion
is an alternating geometric series with expansion coefficient
$1/2$. Assuming that the same pattern is found at higher orders we
estimate the remainder of $O(b^n)$ to be $-1/3$ of the last computed
term and estimate
\begin{equation}
G^{\rm Imp}_{\mu\mu}(0)= 0.1549\,(1+\xi/4) - 0.0266 \pm 0.0001\, .
\end{equation}
The error has been estimated as half of the remainder. We will
encounter the same behavior for the quark self-energy graphs and calculate
our final results also in that case with the same prescription. 

Let us note that Eq.(\ref{eqprop}) allows us to compute the so-called
``mean link'', a quantity needed to perform tadpole improvement
\cite{Lepage:1992xa} of lattice perturbation theory.  At $O(\alpha_s)$
the mean link in Landau gauge $\xi=-1$ is defined as
\begin{equation}\label{u02}
\langle U_\mu(x)/3 \rangle =1+\frac{8\pi\alpha_s}{3}G^{\rm Imp}_{\mu\mu}(0).
\end{equation}.
Using Eq.(\ref{eqprop}) we easily find:
\begin{equation}
\langle U_\mu(x)/3 \rangle = 1+ \alpha_s u_0 
=1+\alpha_s(\,0.7504\pm 0.001)\,,
\end{equation}
in agreement with Ref.~\cite{Nobes:2001tf}.

\section{Quark self-energy}

The one loop quark self-energy has the form
\begin{equation}
\Sigma(p)=\alpha_s\left[i p\,\!\!\!/ \Sigma^{(1)}+ m\,  \Sigma^{(2)}\right]+O(a^2)\,.
\end{equation}

\begin{figure}[htb]
\begin{center}
\includegraphics[height=0.07\textwidth]{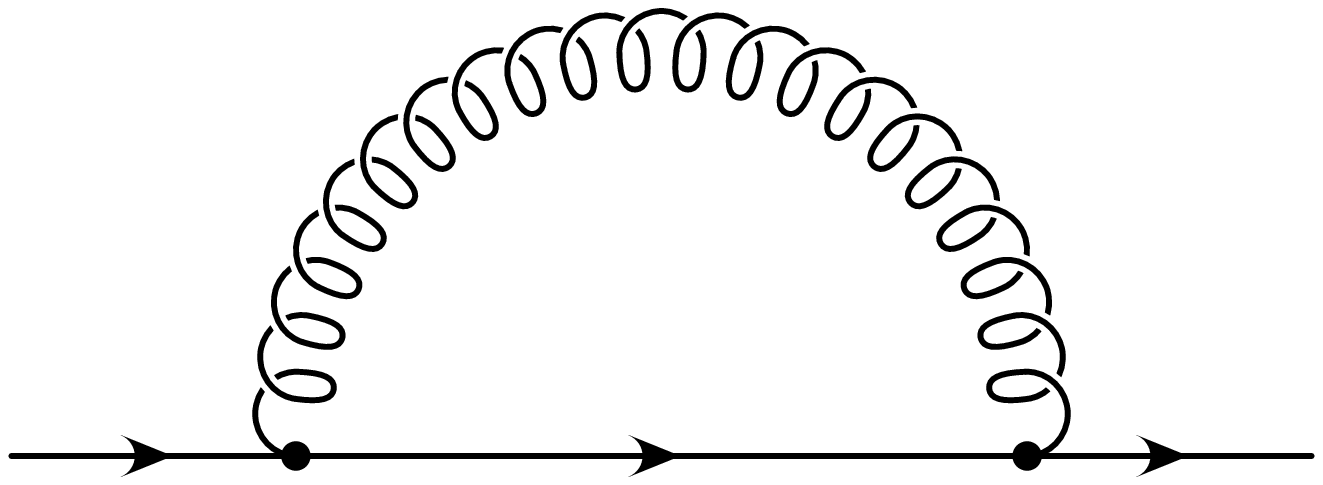}\phantom{abc}\includegraphics[height=0.07\textwidth]{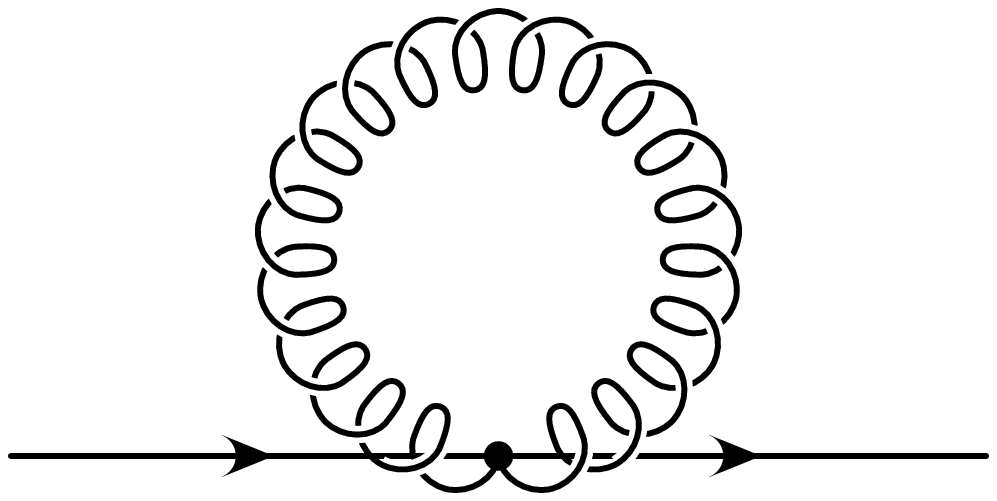}
\end{center}
\vspace*{-0.5cm}
\caption{The two graphs contributing to the fermion self-energy.}
\label{fig:graphs}
\end{figure}

Below, we separately give the contributions of the two graphs in
Fig.~\ref{fig:graphs}. Since the expressions for the graphs in terms
of master integrals are neither short nor particularly illuminating,
we have evaluated them numerically. The parameter $b=1$ indicates the
order in the expansion of the propagator.

\subsubsection*{Soft part}
The leading order soft part is obtained by evaluating the continuum
diagram in analytic regularization. The reason for that is that all
the improvements are subleading and also the fermion doublers decouple
in the continuum limit. The doublers contribution is suppressed
because the appearance of a doubler requires at least one of the gluon
momenta components to be of the order of the lattice spacing.

We therefore get the same result as for Wilson fermions
\begin{eqnarray}
&& \Sigma^{(1)}_{\rm soft}=(1+\xi)\,\Big\{-L_\delta + x - \frac{1}{2} 
+ \left(1 -  x^2 \right) \,\ln\frac{1+x}{x} \Big \} \, , \nonumber\\
&& \Sigma^{(2)}_{\rm soft} =\left ( 4 +
\xi \right) \,\Big\{-L_\delta  
-1 + ( 1+ x)\,\ln\frac{1+x}{x}
 \Big \} \, .\nonumber 
\end{eqnarray}
with $x=m^2/p^2$, $L_\delta=1/\delta - \ln(m^2/4)$.

\subsubsection*{Hard part}

We now compute the hard part of the tadpole and the sunset diagrams. 
As it turns out, the tadpole contribution is roughly an order of 
magnitude larger than that of a sunset; for this reason we have 
to evaluate more terms in the expansion for the tadpole than for 
the sunset. We obtain
\begin{eqnarray}
\Sigma^{(1)}_{\rm hard} &=& 2.6524-1.4214\,b+0.5334\,b^2
\\ \nonumber && 
-0.2452\,b^3+0.1233\,b^4 
\\ \nonumber &&
-0.0655\,b^5+(0.3245-0.0345\,b)\,\xi\\ \nonumber
&=& 1.60(1) + 0.290\,\xi , \\
 \Sigma^{(2)}_{\rm hard} &=&0 .\nonumber 
\end{eqnarray}

For the hard part of the sunset diagram we find:
\begin{eqnarray}
\Sigma^{(1)}_{\rm hard}& =&\frac{1+\xi}{3\pi \delta} 
-0.8327+0.0880\,b-0.0183\,b^2 \nonumber \\
&& -(0.6328-0.345\,b)\,\xi , \nonumber \\
\Sigma^{(2)}_{\rm hard} 
&=&\frac{4+\xi}{3\pi \delta}-0.9821+0.3279\,b-0.0857\,b^2.
\end{eqnarray}

\subsubsection*{Pole mass}

It is straightforward to extract the pole mass from the above formulas.
The pole mass is defined as the position of the pole of the fermion 
propagator. It 
therefore reads:
\begin{eqnarray}
m_{\rm pole}&=&=m_{\rm bare}\,(1+\alpha_s\,\Delta_1),\nonumber\\ 
\Delta_1 &=&\Sigma^{(1)}-\Sigma^{(2)}\nonumber\\
&=&-\frac{2}{\pi}\ln m + 1.60(1) + 0.66(1) \,.
\label{eqmass1}
\end{eqnarray}
The first number is the contribution of the tadpole graph, while the
second one corresponds to the sunset diagram.  Note that the hard and the 
soft contributions to the pole mass are not gauge
invariant separately. 
This is the consequence of putting the regulator $\delta$ 
on the fermion propagator, which we are forced to do to properly deal 
with the doublers. The gauge dependence only cancels in the final 
result for the pole mass, when the hard and the soft contributions 
are added together.

The effect of the improvement terms in the lattice action is
large. This can be seen by comparing with the result obtained when all
the improvements are switched off. In that case the relation between
the pole and bare masses changes to
\begin{eqnarray}
m_{\rm pole}^{\rm naive}&=&=m_{\rm bare}\,(1+\alpha_s\,\Delta_1^{\rm naive}),\nonumber\\ 
\Delta_1^{\rm naive} &=&-\frac{2}{\pi}\ln m +  4.586 \,.
\end{eqnarray}
One sees that the renormalization constant in this case is roughly a factor of two larger than for the improved action. The reduction arises because the
smearing of the link in Eq.~(\ref{Vlink}) strongly reduces the quark gluon
interaction,  while the effect
of all other improvements is significantly smaller.

Note that the result in Eq.(\ref{eqmass1})
is not ``tadpole-improved'' \cite{Lepage:1992xa}. 
For a tadpole improved action, 
an additional tree level contribution
\begin{equation}
\Delta^{\rm tad}= -\left(\frac{5}{2}-\frac{b}{4}\right)\, \alpha_s\,u_0^{(2)}
\end{equation}
arises, which tends to cancel the tadpole graph.  The quantity
$u_0^{(2)}$ is the $O(\alpha_s)$ coefficient of the mean link in Landau
gauge, given in Eq.(\ref{u02}) so that 
\begin{equation}
\Delta^{\rm tad}=-1.688(2).
\end{equation}
Putting everything together, restoring the lattice 
spacing and combining the numerical errors in  
quadratures, we derive the following relation 
between the pole and the bare quark mass 
in the continuum limit:
\begin{equation}
m_{\rm pole} = m_{\rm bare} \left [ 1 + \alpha_s 
\left ( -\frac{2}{\pi} \ln ma +0.572(14) \right) \right ].
\label{eqres}
\end{equation}
Our results agree with the findings of Ref.~\cite{Hein:2001kw}. Note that this
reference uses the mean field $u_0$ defined from the average plaquette, for which $u_0^{(2)}\approx 0.7671$ \cite{Nobes:2001tf} .

\section{Conclusion}

Perturbative calculations with lattice regularization are required to
provide a bridge between non-perturbative simulations and the
continuum physics. In this paper, we have described an analytic
calculation of the fermion self-energy of improved staggered fermions and the L\"uscher-Weisz action for
gluons.  This action has been recently used for non-perturbative
lattice simulations by the MILC collaboration.

Our main result, the relation between the pole and the bare lattice
quark masses is given in Eq.(\ref{eqres}). We observe that the
renormalization constant for the improved action is rather
modest. This is mainly a consequence of the smearing of the
quark-gluon interaction that was introduced to strongly suppress the
interaction between the fermions in different corners of the Brillouin
zone.  Let us stress that in the current calculation the doublers do
not contribute in the continuum limit even for the naive action; the
strong change in the renormalization constant therefore has not so
much to do with their decoupling as with a significant reduction of
the quark-gluon interaction interaction over the entire Brillouin
zone.

This calculation demonstrates that algebraic techniques for 
lattice perturbation theory, based  on the methods developed 
for continuum perturbation theory, are quite efficient. 
With the large database for one-loop lattice tadpole integrals 
for staggered fermions,  which we have produced in the 
course of this calculations, other one-loop
perturbative lattice calculations for improved lattice actions 
seem to be within reach.

{\bf Acknowledgments} We are grateful to Q.~Mason for providing the
Feynman rules for the improved action. We thank C.~Anastasiou for
letting us use his computer program \cite{babis} to perform the
reduction of to master integrals. This research was supported in part
by the DOE grants DE-AC03-76SF00515 and DE-FG03-94ER-40833.


\begin{thebibliography}{9}


\bibitem{Becher:2002if}
T.~Becher and K.~Melnikov,
Phys.\ Rev.\ D {\bf 66}, 074508 (2002).

\bibitem{Smirnov:pj} For a review, see
V.~A.~Smirnov, {\it ``Applied Asymptotic Expansions In Momenta And Masses,''}
Berlin, Germany: Springer (2002) 262 p.

\bibitem{Bernard:2001av}
C.~W.~Bernard {\it et al.},
Phys.\ Rev.\ D {\bf 64}, 054506 (2001)
[arXiv:hep-lat/0104002].

\bibitem{Luscher:1985zq}
M.~L\"uscher and P.~Weisz,
Phys.\ Lett.\ B {\bf 158}, 250 (1985).

\bibitem{Kogut:ag}
J.~B.~Kogut and L.~Susskind,
Phys.\ Rev.\ D {\bf 11}, 395 (1975).


\bibitem{Lepage:1998vj}
G.~P.~Lepage,
Phys.\ Rev.\ D {\bf 59}, 074502 (1999)
[arXiv:hep-lat/9809157].


\bibitem{Nobes:2001tf}
M.~A.~Nobes, H.~D.~Trottier, G.~P.~Lepage and Q.~Mason,
Nucl.\ Phys.\ Proc.\ Suppl.\  {\bf 106}, 838 (2002)
[arXiv:hep-lat/0110051].



\bibitem{Lepage:1992xa}
G.~P.~Lepage and P.~B.~Mackenzie,
Phys.\ Rev.\ D {\bf 48}, 2250 (1993)
[arXiv:hep-lat/9209022].


\bibitem{Hein:2001kw}
J.~Hein, Q.~Mason, G.~P.~Lepage and H.~Trottier,
Nucl.\ Phys.\ Proc.\ Suppl.\  {\bf 106}, 236 (2002)
[arXiv:hep-lat/0110045].

\bibitem{babis} C.~Anastasiou, to be published.




\end{thebibliography}
\end{document}